\documentclass [10pt]{article}

\usepackage[makeroom]{cancel}

\usepackage{amsmath}

\usepackage{slashed}

\usepackage{hyperref}
\usepackage{geometry}

\usepackage{graphicx}
\usepackage{amssymb}
\usepackage{epstopdf}

\usepackage{cite}
\usepackage{graphicx}
\usepackage[usenames]{color}
\usepackage{subfigure}
\usepackage{setspace}
\usepackage{slashed}
\usepackage{multirow}
\usepackage{ulem}

\usepackage[hang,small]{caption}

\usepackage{geometry}
    \usepackage{amsmath}   
        \usepackage{amssymb}   
    
    \newcommand{\ignore}[1]{}
 \newcommand{\ee}{\end{equation}}

\def\ba#1\ea{\begin{align}#1\end{align}}
\newcommand{\bit}{\begin{itemize}}
\newcommand{\eit}{\end{itemize}}

\newcommand{\nn}{\nonumber} \renewcommand{\bf}{\textbf}

\newcommand{\e}{\mathrm{e}} 
  \newcommand{\g}{\gamma}

\usepackage{fancyhdr}
\newcommand{\iu}{{i\mkern1mu}}
\usepackage{titlesec,titletoc}
  \titleformat{\section}{\Large\sf\bfseries}{\thesection}{1em}{}
  \titleformat{\subsection}{\large\sf\bfseries}{\thesubsection}{1em}{}

\title{\sf\bfseries \ntitle}
\author{
    Sumeet Dagaonkar\footnote{sumeetkd@iitk.ac.in}~, Pankaj Jain\footnote{pkjain@iitk.ac.in}~\\
    \it{Department of Physics, Indian Institute of Technology, Kanpur 208016, India}\\
\and 
John P. Ralston\footnote{ralston@ku.edu}\\
\it{Department of Physics \& Astronomy, University of Kansas,}\\
\it{Lawrence, KS - 66045, USA}\\
}

\date{}%{\today}

\newcommand{\pghdr}{\footnotesize {S. Dagaonkar} {\it et al.} -- The Dirac Form Factor Predicts \dots }
\newcommand{\ntitle}{The Dirac Form Factor \\ Predicts the Pauli Form Factor \\ in the Endpoint Model}

\begin{document}

\vspace{-3cm}

\maketitle
\begin{abstract}{
We compute the momentum-transfer dependence of the proton Pauli form factor $F_{2}$ in the endpoint
overlap model. We find the model correctly reproduces the scaling of the ratio of $F_{2}$ with the
Dirac Form factor $F_{1}$ observed at the Jefferson Laboratory. The calculation uses the
leading-power, leading twist Dirac structure of the quark light-cone wave function, and the same
endpoint dependence previously determined from the Dirac form factor $F_{1}$. There are no
parameters and no adjustable functions in the endpoint model's prediction for $F_{2}$. The model's
predicted ratio $F_{2}(Q^{2})/F_{1}(Q^{2})$ is quite insensitive to the endpoint wave function,
which explains why the observed ratio scales like $1/Q$ down to rather low momentum transfers. The endpoint
model appears to be the only comprehensive model consistent with all form factor information as well
as reproducing fixed-angle proton-proton scattering at large momentum transfer. Any one of the
processes is capable of predicting the others.  } 
\end{abstract}
\vspace{1cm}

\section{Introduction}
The electromagnetic form factors know as $F_{1}$ and $F_{2}$ are an important probe of the internal
structure of nucleons. A popular theoretical model assumes that 
at high momentum transfer these quantities can be 
factorized into a hard scattering contribution and a so-called distribution amplitude. 
The distribution amplitude has no information about the
proton wave function except the parton momentum fraction Feynman-$x$ dependence and some spin factors of a
short-distance expansion.  The focus of the short-distance (SD) model 
\cite{Brodsky:1974vy,Farrar:1979aw,Lepage:1980fj,Radyushkin80,Radyushkin80a,Brodsky:1981kj} is a perturbatively calculable hard scattering kernel. The model generates an order
by order expansion in powers of the inverse momentum transfer-squared, $1/Q^2$. The expansion has
often been claimed to be the unique prediction of QCD. However the task of comparing the model to
the larger theory of QCD was never completed, and obviously cannot be explored within the SD model
itself. 

Yet model predictions can be compared to experimental data. The SD model predicts that
$F_{2}(Q^{2})/F_{1}(Q^{2}) \rightarrow 1/Q^2 $ for large $Q$ \cite{Brodsky:1974vy, Ji:1986um}. A
simple way to obtain this uses a perturbative quark mass to flip two quark helicities in the
internal lines. 
The experimental results obtained at the Jefferson lab\cite{Jones:1999rz,jlab-gayou}, however,
showed that the ratio $F_{2}(Q^{2})/F_{1}(Q^{2}) \sim 1/Q$ in the energy range $2\ {\rm
GeV}^{2}<Q^{2}<5.6\ {\rm GeV}^{2}$. This contradicted the prediction of the SD model that some
thought had been established. The results played an important role in dramatizing the failure of the
SD model, which had also been anticipated earlier \cite{Isgur,JPR}. It is now clear that the SD model might apply only at very large energies, which are
inaccessible experimentally. Even at asymptotic energies there is no proof the model dominates.

Since the SD model fails it is imperative to explore alternatives.

Work by Miller et al, Lin et al,
and Cloet et al \cite{Miller:2002qb,Lin:2010fv,Cloet:2014rja} has reproduced the experimentally
observed momentum dependence of $F_2$. These calculations emphasize the importance of the quark wave
functions, i.e. the role of hadron structure, as opposed to the role of perturbation theory. 

Kivel and Vanderhaeghen\cite{Kivel:2010ns} used Soft Collinear 
Effective Theory (SCET) to analyze
soft spectator contributions which represent a class of diagrams which also give the SD scaling of
$Q^2 F_{2}/F_{1} \sim const.$ for large $Q^{2}$. 
 It is possible that at smaller $Q^2$ of order few GeV$^2$, 
the soft spectator contributions might lead to the observed
experimental behavior. 
However, it is not clear how to extract these contributions systematically.
In an earlier analysis, 
Belitsky et.al  \cite{Belitsky:2002kj} obtained the dependence $F_{2}(Q^{2})/F_{1}(Q^{2}) \rightarrow
1/Q^2*\log(Q^{2}/\Lambda_{QCD}^{2}) $, which matches with the observed data, by
introducing higher twist light cone wave functions. The 
logarithmic term is a result of an
integration over the soft endpoint region, where the assumptions of the 
SD model no longer hold. Hence we find that
some studies in the past \cite{Kivel:2010ns,Belitsky:2002kj} have attributed the observed  
experimental behavior of $F_2/F_1$ to the contributions arising from the
soft spectator quarks in the end point region.

The significance of the end point region for the calculation of the
ratio $F_2/F_1$ is rather interesting in view of the recent claim  
 that an end point model 
  ($EP$ model) can comprehensively
explain the scaling behavior of many exclusive processes \cite{Brodsky:1974vy,matveev,Sivers76}. The model relates the observed scaling to
the behavior of the quark wave function as Feynman-$x\rightarrow 1$. In this limit one of the quarks
carries most of the proton longitudinal momentum. The model appeared several times in the
literature, yet it was dismissed prematurely, often for reasons that its premises contradicted the
assumptions of the SD model. For that reason the $EP$ region was long regarded as a nuisance. Many
efforts attempted to show the $EP$ model's contribution would be suppressed, but the efforts were
unsuccessful.

Once given fair consideration, the $EP$ model appears to provide the simplest explanation of several
experimental observations. In \cite{Dagaonkar:2014}, we applied the $EP$ model to compute the pion
form factor,  the proton Dirac form factor $F_{1}(Q^{2})$, and the proton-proton elastic scattering
cross section at high momentum transfer. We found that one consistent wave function for the end
point region could be extracted by fitting the experimental form factor data. The same wave function
then predicts the scaling behavior observed in proton-proton fixed-angle scattering.
We extend this study here in order to determine the proton Pauli
form factor $F_{2}$. We find that the formalism predicts
$F_{2}(Q^{2})$ without introducing any new parameters.  

Let us briefly explain the physics. It is well known that quark mass insertions produce a quark
helicity flip, which can ultimately produce the proton helicity (more specifically, chirality) flip
characterizing $F_2 $. Quark mass terms are negligible in the high energy limit of the SD model.
This is because they compete with terms scaling like the large momentum $Q$. The role of a quark
mass is qualitatively different in the $EP$ model.  The soft quarks with momentum fractions $x\sim
0$ already have very small momenta. Their momenta are of the order of the QCD chiral symmetry
breaking scale $\Lambda$. Then an equally small contribution from a quark mass is not a relatively
small effect, and it cannot be neglected.  Let us repeat that the attempt to banish small momentum
regions from QCD never worked out. An unexpected consequence of small momenta appearing at leading
power order is that mass effects can appear at the same order.

Another effect makes this even more interesting. Under a Lorentz transformation with rapidity $y$ in
the $z$ direction, the big light cone + component transforms by $e^y$ and the small component
like $e^{-y}$. All previous calculations known to us at leading power order integrate quark wave
functions over the small momentum in the first step. This appears to be much more safe than
integrating over the transverse momentum components, which scale like $1$ compared to $e^{-y}$.  Yet
we have discovered a limit-interchange error occurs. Integrating away the small components is the
first step of the SD model producing a visible factorization into separated hadronic parts. The
assumption, actually a hope, that some factorization dominates is what demands that step. Yet that
step instantly causes $F_2$ to scale no larger than $1/Q^6$. When the small momenta components are
retained  in the scattering process we find a contribution to $F_2$ scaling like $1/Q^5$.  The
integrals cannot be represented by effective, pre-integrated quantities that depend only on
Feynman-$x$.  This phenomenon contradicts the tenets of factorization. 
In the $EP$ model, the leading
power contribution to $F_2$ comes from an inseparable union of initial and final state proton
states. 

Finally all of this occurs with one simple wave function, which happens to be the most often
cited, leading twist example. There is no particular reason to favor leading twist coming from a
short distance expansion. There is every reason to use a wave function of leading power in the large
momentum $P$. It is seldom noticed that the leading power, leading twist wave function has both
chirally-even and chirally-odd components. A single 
wave function can both maintain the proton's
chirality in $F_{1}$, and flip the chirality in $F_{2}$. 

In section \ref{sec:EP}, we show that by respecting the necessary integration region, while using
the endpoint dependence of the proton wave function obtained in \cite{Dagaonkar:2014}, we obtain the
experimentally observed scaling behavior for $F_2/F_1$. This is a remarkable prediction of the model: If
attention had been given 30 years ago, it would have predicted $F_2$ in advance of the data.
Reversing the argument, the observed scaling dependence of $F_2/F_1$ predicts $F_1$ and $pp$ scattering
at high momentum transfer. None of these facts requires appealing to an unusually large logarithmic
correction, or an unusually large dimensionful scale. As far as we know it is the first time that
one model is actually consistent with the known data. 

Quark orbital angular momentum is a topic of great interest.  No orbital angular momentum ($OAM$)
enters the SD model, because a theoretical preference for factorization demands integrating over
quark transverse momenta before the actual reaction has even been set up.  Information about
transverse size is lost by that step.  When $OAM$ is re-cast into a twist expansion
\cite{Belitsky:2002kj} the sequence of operations dictated by the SD model produces a $1/Q^6$
dependence for $F_2$.  References \cite{Ralston:2003mt,Jain:2003za} showed that avoiding the SD
assumptions and performing the transverse momentum integrations  to compute $F_2$ led to power law
dependence for $F_2 $ intermediate between $1/Q^4$ and $1/Q^6$. That is, the integration region
assumed to dominate asymptotically was not the actually dominant region, whether or not an endpoint
issue was considered. While the asymmetry of the endpoint integration regions produces a rather
obvious role for $OAM$, that is not the focus of this paper. This paper is about using the same
leading twist Dirac and endpoint structure found in the $F_{1}$ calculation to calculate $F_{2}$.
The calculation is relatively simple, and agrees remarkably with data. Even more remarkably, the
ratio $F_{2}(Q^{2})/F_{1}(Q^{2})$ is quite insensitive to the endpoint wave function, explaining why
the observed ratio goes like $1/Q$ down to rather small momentum transfer. 

\section{Endpoint Calculation}
\label{sec:EP}

\begin{figure}[htbp]
\begin{center}
\includegraphics[width=4in]{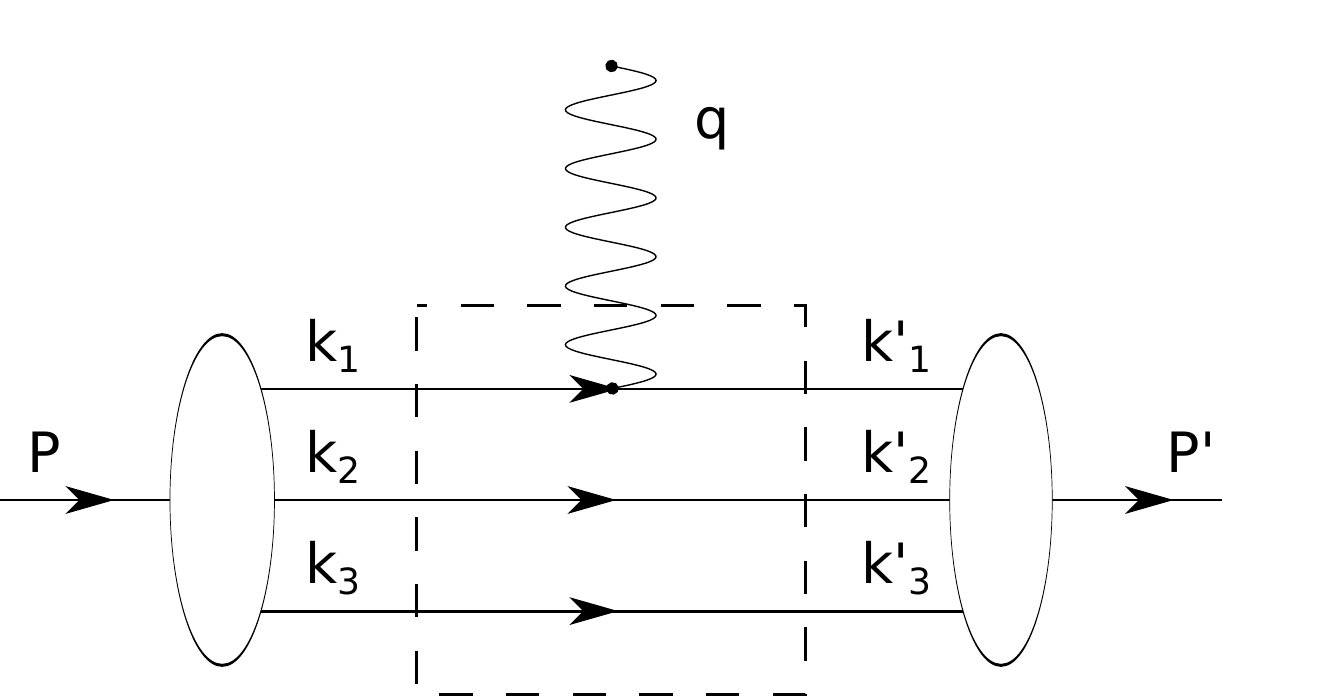}
\caption {\small The basic kinematics of the end point contribution
to the proton form factor. The photon with momentum $q$ scatters with one of
the quarks which carries the dominant fraction of the proton momentum. }
\label{fig:gammapp}
\end{center}
\end{figure}
Here we describe the calculation of $F_2$ through the quark mass contribution.  One quark is struck by the virtual photon. The remaining quarks will be in a small momentum region, such that their incoming and outgoing momenta are entirely determined by their wave functions.  No interactions are computed for those particles, because perturbative interactions would double-count what is already included in the wave functions.
We will use the same wave functions to compute $F_2$ as previously determined \cite{Dagaonkar:2014} from $F_1$, found to be consistent with $pp$ scattering.

\subsection{Coordinates} 

The basic diagram   for proton electromagnetic form factor is given by Fig. \ref{fig:gammapp}.  The initial and final proton 4-momenta are $ P$ and $ PÕ$, with $q=PÕ-P$.  Initial quark momenta $k_j$ (masses $m_j$) are unprimed, while final momenta use the same label with a prime.  We let $k_{1}$  denote the struck quark, and $k_2$, $k_3$ denote the spectators.  
Our coordinates are $(energy, \,p_{x}, \, p_y, \, p_z).$ We use a Lorentz frame where the incoming and outgoing protons momenta are 
\ba
P^{\mu} &= \left(\sqrt{{Q^{2}\over 2} + m_{P}^2},-{Q \over 2},0 ,{Q \over 2}
\right), \nn \\
P'^{\mu} &= \left(\sqrt{{Q^{2}\over 2} + m_{P}^2},{Q \over 2},0 ,{Q \over 2}\right), \nn \\
q^{\mu} &= (0,Q,0,0)\,.
\ea
Here $m_{P}$ is the mass of the proton.

We introduce a basis for transverse momenta: 
\ba y^{\mu} =&(0, \, 0, \, 1, \,0) = y'; \quad \hat P\cdot  y=\hat P'\cdot 
 y'=0; \nn \\ n^{\mu} =&\frac{1}{\sqrt{2}} (0, \, -1, \, 0, \, -1), \quad \hat P\cdot n= 0;\nn \\  n'^{\mu} =&\frac{1}{\sqrt{2}} (0, \, 1, \, 0, \,-1),
\quad \hat P'\cdot n'=0. \nn 
\ea
Here $\hat P = (0,-\frac{1}{\sqrt{2}},0,\frac{1}{\sqrt{2}})$ and $\hat P' = (0,\frac{1}{\sqrt{2}},0,\frac{1}{\sqrt{2}})$ are the unit vectors along
the direction of propagation of the incoming and outgoing protons respectively.
The components of the quark 3-momenta are
expressed as
 \ba \vec{k}_{i}& = x_{i}\frac{Q}{\sqrt{2}}\hat{P}+ k_{in}\vec n +k_{iy} \vec y =(-x_{i}Q/2, \, 0, \, x_{i}Q/2) + (-k_{in}/\sqrt{2}, \, k_{iy}, \, -k_{in}/\sqrt{2}); \nn \\ 
 \vec{k}^{\,'}_{i}& =x'_{i}\frac{Q}{\sqrt{2}}\hat{P}^{'}+ k'_{in} \vec n'+ k'_{iy} \vec y= (x'_{i}Q/2, \, 0, \, x'_{i}Q/2  ) + (k'_{in}/\sqrt{2}, \, k'_{iy}, \, -k'_{in}/\sqrt{2}).\label{eq:qmomenta}
\ea
The four momenta of the quarks are then given by,
\ba
k_{i}^{\mu} &=   \left(k_{i}^{0},-x_{i}\frac{Q}{2}-\frac{k_{in}}{\sqrt{2}},k_{iy},x_{i}\frac{Q}{2}-\frac{k_{in}}{\sqrt{2}}\right) \nn\\
k_{i}^{'\mu} &= \left(k_{i}^{'0},x'_{i}\frac{Q}{2}+\frac{k'_{in}}{\sqrt{2}},k'_{iy},x'_{i}\frac{Q}{2}-\frac{k'_{in}}{\sqrt{2}}\right).
\label{eq:qmomentum}
\ea

\subsection{The Matrix element} 

With $ J^{\mu}$ the electromagnetic current operator and $N$ standing for Dirac spinors, the matrix element for the interaction is parameterized by
\ba <p^\prime s^\prime|J^{\mu}|ps> = -\iu e \left[F_{1}(Q^{2})(\overline{N}^{\prime}\gamma^{\mu}N) +
\frac{ F_{2}}{2 m_{p}}(Q^{2}) \overline{N}^{\,\prime} \iu\sigma^{\mu\nu}q_{\nu}N \right]\, \ea

Let $\Psi_{\alpha\beta\gamma}$ be the Bethe-Salpeter 3-quark wave function in the proton with spinor
indices shown. Let symbol $\mathcal{M}^{\mu}$ stand for the quark-photon vertex, propagator factors,
and momentum conservation factors, displayed in a moment. The model for the reaction is 
\ba
<p^\prime s^\prime|J^{\mu}|ps> = &\int \prod_{i}\frac{d^{4}k_{i}}{(2\pi)^{4}}
\frac{d^{4}k'_{i}}{(2\pi)^{4}}\delta^{4}(k_{1}+k_{2}+k_{3}-P)\delta^{4}(k'_{1}+k'_{2}+k'_{3}-P') \nn
\\ & \times \left[ \overline{\Psi}_{\alpha'\beta'\gamma'}^{\prime}(k^{'}_{i}) \times
\mathcal{M}_{\alpha'\beta'\gamma'\alpha\beta\gamma}^{\mu}\times \Psi_{\alpha\beta\gamma}(k_{i})
\right]  \label{eq:formy} 
\ea
 
Here $\mathcal{M}^{\mu}$ is 
\ba
\mathcal{M}^{\mu}=&-ie\gamma^{\mu}_{\alpha^{'}\alpha}\delta^{4}(k_{1}+q-k'_{1})(\slashed{k}_{2}-m_{2})_{\beta^{'}\beta}\delta^{4}(k_{2}-k'_{2})(\slashed{k}_{3}-m_{3})_{\gamma^{'}\gamma}\delta^{4}(k_{3}-k'_{3})  \nn \\
      &-ie\gamma^{\mu}_{\beta^{'}\beta}\delta^{4}(k_{1}+q-k'_{1})(\slashed{k}_{2}-m_{2})_{\gamma^{'}\gamma}\delta^{4}(k_{2}-k'_{2})(\slashed{k}_{3}-m_{3})_{\alpha^{'}\alpha}\delta^{4}(k_{3}-k'_{3})  \nn\\
      &-ie\gamma^{\mu}_{\gamma^{'}\gamma}\delta^{4}(k_{1}+q-k'_{1})(\slashed{k}_{2}-m_{2})_{\alpha^{'}\alpha}\delta^{4}(k_{2}-k'_{2})(\slashed{k}_{3}-m_{3})_{\beta^{'}\beta}\delta^{4}(k_{3}-k'_{3}). \label{eq:m} 
\ea

Note the delta functions $\delta^{4}(k_{2}-k'_{2})\delta^{4}(k_{3}-k'_{3})$ which explicitly enforce momentum conservation of spectator quarks.

The initial light cone coordinates are defined as
\ba 
\nn k^{+}_{i}=k^{0}_{i}+\frac{x_{i}Q}{\sqrt{2}}; \qquad  k^{-}_{i}=k^{0}_{i}-\frac{x_{i}Q}{\sqrt{2}} . \nn \ea Final state symbols have a prime.
In literature, it is standard to use the co-ordinates $\kappa^{-} = k^{-}p^{+}$; $\kappa^{+} = k^{+}/p^{+}$;
$\kappa^{'-}= k^{'-}p^{'+};\kappa^{'+}= k^{'+}/p^{'+}$ which is just parameterizing the light cone
co-ordinates with the momenta $p^{+},p^{'+}$.

\subsection{Integration} 
It is generally assumed that wave functions $ \overline{\Psi^{\prime}}(k'_{i})$, $\Psi(k_{i}) $ of
4-momenta are peaked near the on-shell region. In that region, the actual wave function can be
replaced by its integral over the small momentum component $\kappa_{i}^{-}, \kappa_{i}^{'-}$,
producing the usual light cone wave function $ \overline{Y^{\prime}}(x'_{i},\vec{k}'_{\perp i}),
Y(x_{i},\vec{k}_{\perp i})$\cite{Brodsky:1984vp}.
 
\ba
 & <p^\prime s^\prime|J^{\mu}|ps> =\int \prod_{i}\frac{d\kappa^{+}_{i}d\vec{k}_{\perp i}}{(2\pi)^{4}}\frac{d\kappa^{'+}_{i}d\vec{k}'_{\perp i}}{(2\pi)^{4}}
\delta(\kappa^{+}_{1}+\kappa^{+}_{2}+\kappa^{+}_{3}-1)
\delta^{2}(k_{\perp 1}+k_{\perp 2}+k_{\perp 3})\\ \nn
&\delta(\kappa^{'+}_{1}+\kappa^{'+}_{2}+\kappa^{'+}_{3}-1) \delta^{2}(k'_{\perp 1}+k'_{\perp 2}+k'_{\perp 3}) 
 (\int \prod_{j}d\kappa^{-}_{j}\delta(\kappa^{\prime-}_{1}+\kappa^{\prime-}_{2}+\kappa^{\prime-}_{3}-m^{2}_{p})  \overline{\Psi^{\prime}}_{\alpha^{'}\beta^{'}\gamma^{'}}(k^{'\mu}_{i}))
 \\&\nn \mathcal{M}^{\mu}(k_{i},k^{\prime}_{i})  (\int \prod_{l} d\kappa^{-}_{l}\delta(\kappa^{-}_{1}+\kappa^{-}_{2}+\kappa^{-}_{3}-m^{2}_{p})\Psi_{\alpha\beta\gamma}(k^{\mu}_{i})) 
\ea
The rest of the calculation cannot use the same approximation, because the delta-functions vary
rapidly: Hence the process is indivisibly linked together by the integrations.  

The above expression for scattering kernel $\mathcal{M}^\mu$ has an important dependence on $\kappa_2^-$ and
$\kappa_3^-$, which cannot be overlooked. This is the point where our calculation begins to differ
from previous ones. 

The basic problem is that for the soft spectator quarks it is not reasonable
to assume that their four momentum square, $k^2$, is approximately zero. 
We expect $k^2$ to be of the order of $\Lambda^2$. In a constituent quark
model, these quarks are assumed to be approximately 
on-mass-shell with masses of the 
order of few hundred MeV for the up and down quarks. In general the behavior
of the quark propagator is expected to be more complicated and one can
model its form by solving a truncated Schwinger-Dyson equation \cite{jainmunczek,Alkofer,robertswilliams}. 
For our purpose, 
it is adequate and self-consistent to assume that $\mathcal{M}^{\mu}$ is dominated by the on-shell
region and the $\kappa^{-},\kappa^{'-}$ dependence can be replaced by the on-shell expression,   
\ba \kappa^{-} = \frac{m_{q}^{2}+\vec{k}_{\perp}^{2}}{\kappa^{+}} \ea

As explained above, we assume that the mass of the slow spectator quarks is
 of the order of a few hundred MeV. 
 On the other hand, the struck quark is a
perturbative object. Treating it consistently uses a mass of order of a few MeV. We ignore the tiny
and power-suppressed helicity-flip contributions from the struck quark.

Now, doing a change of variables gives the standard form with 

\ba
<p^\prime s^\prime|J^{\mu}|ps> = \int \prod_{i} \frac{dx_{i}d\vec{k}_{\perp i}}{(2\pi)^{3}} \frac{dx^{'}_{i}d\vec{k}'_{\perp
i}}{(2\pi)^{3}} & \delta(x_{1}+x_{2}+x_{3}-1)\delta^{2}(k_{\perp 1}+k_{\perp 2}+k_{\perp
3})\delta(x'_{1}+x'_{2}+x'_{3}-1)\nn\\
& \delta^{2}(k'_{\perp 1}+k'_{\perp 2}+k'_{\perp 3})
\bigg( \overline{Y^{\prime}}_{\alpha^{'}\beta^{'}\gamma^{'}}(x'_{i},\vec{k}'_{\perp
i}) \mathcal{M}^{\mu}  Y_{\alpha\beta\gamma}(x_{i},\vec{k}_{\perp i}) \bigg) 
     \label{eq:formfinal}
\ea

The delta functions of Eq.(\ref{eq:formfinal}) and Eq.(\ref{eq:m}) lead to the following conditions,
\ba
\nn & x_{3} = 1 - x_{1} - x_{2};\,\quad x'_{3} = 1 - x'_{1} - x'_{2};\\ \nn & k_{1n} = -k_{2n} -
k_{3n}; \quad k_{1y} = -k_{2y} - k_{3y} ; \nn \\ 
&  k'_{1n} = -k'_{2n} - k'_{3n}; \quad k'_{1y} = -k'_{2y} - k'_{3y} ;\nn \\ 
& k_{1y} = k'_{1y}; k_{2y} = k'_{2y} ;\,\quad x'_{1}= x_{1}+
\mathcal{O}\left(\frac{\Lambda}{Q}\right); x'_{2}= x_{2}+ \mathcal{O}\left(\frac{\Lambda}{Q}\right);\nn\\ 
& k_{1n}=\frac{Q}{\sqrt{2}}(1-x'_{1}) ;\,\quad k'_{1n}=\frac{Q}{\sqrt{2}}(1-x_{1}); \nn\\
& k_{2n}=\frac{Q}{\sqrt{2}}(-x'_{2}) ;\,\quad k'_{2n}=\frac{Q}{\sqrt{2}}(-x_{2}). \nn
\ea

The light cone wave function $Y$ of leading twist and leading power of large $P$ is
\cite{Ioffe,Avdeenko},
 \begin{equation}      Y_{\alpha\beta\gamma}(k_{i},P) = \frac{f_{N}}{16\sqrt{2}N_{c}}\{ 
     (\slashed{P}C)_{\alpha\beta}(\gamma_{5}N)_{\gamma}\mathcal{V} + (\slashed{P}\gamma_{5}C)_{\alpha\beta}N_{\gamma}\mathcal{A} + \iu (\sigma_{\mu\nu}P^{\nu}C)_{\alpha\beta}
 (\gamma^{\mu}\gamma_{5}N)_{\gamma}\mathcal{T}\}.\label{eq:lipwavef}
\end{equation}
 Here $\mathcal{V,A,T}$ are scalar functions of the quark momenta,
         $N$ is the proton spinor,
        $N_{c}$ the number of colors,
        $C$ the charge conjugation operator, 
        $\sigma_{\mu\nu}= \frac{\iu}{2}[\gamma_{\mu},\gamma_{\nu}]$,
and $f_{N}$ is a normalization. This wave function was previously used to compute $F_{1}$, and is now being applied to
 compute $F_2$.

\begin{figure}[ht]
\begin{center}
\includegraphics[width=3.5in]{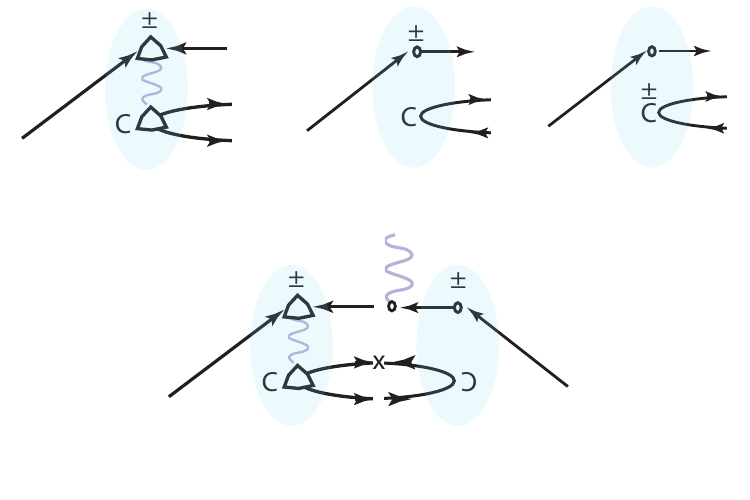}
\caption{ Chirality flow in the calculation of $F_{2}$, indicated by arrows. Chirally-even vertices conserve helicity and chirally odd ones flip helicity. The standard leading twist wave function contains both types, shown across the top. A typical combination of diagrams flipping the final state proton chirality is shown at the bottom. This diagram needs one (1) internal flip of low momentum spectator quark chirality, which appears as the closed loop with a mass insertion indicated by ``X.''}
\label{fig:Ioffe}
\end{center}
\end{figure}

It may come as a surprise that the same chirality structure creating $F_{1}$ can predict $F_{2}$.
Fig. \ref{fig:Ioffe} shows a cartoon of the chirality flow. Each term in the $Y_{\alpha\beta\gamma}$
collection has been classified as {\it chirally even} or {\it chirally odd} depending on whether it
conserves helicity (even, anti-commutes with $\gamma_{5}$) or flips helicity (odd, commutes with
$\gamma_{5})$. Since momentum conservation is trivial it is not shown. The chirality flow of the
$\mathcal{V, A, T}$ terms are shown at the top. A typical combination of diagrams flipping the final
state proton chirality is shown at the bottom. This diagram needs one (1) internal flip of low
momentum spectator quark chirality, which appears as the closed loop with a mass insertion indicated
by ``X.'' The cartoon shows how the Dirac algebra works without needing to do the algebra.  

Returning to Eq. \ref{eq:formfinal}, inserting the wave function Eq. \ref{eq:lipwavef}, and extracting the terms which lead to $F_{2}$ yields
\ba & \overline{N}^{\prime}\frac{\iu}{2}\sigma^{\mu\nu}q_{\nu}N \, F_{2} = \nn  
\int  dk_{1y}dk_{2y}dx_{1}dx_{2}\frac{1}{Q^{2}}\times \big\{ \nn\\
    &  [\overline{N}'\g_{5}\g_{\sigma}\g^{\mu}\g_{5}N] \iu (C^{-1}\sigma^{\rho\sigma}P'_{\rho})_{\alpha'\beta'}(\slashed{k}_{2}-m_{2})_{\alpha'\alpha} (\slashed{k}_{3}-m_{3})_{\beta\beta'}(\slashed{P}C)_{\alpha\beta}\mathcal{V}^{*}\mathcal{T}\nn \\ 
     &+ [\overline{N}'\g_{5}\g^{\mu}\g^{\sigma}\g_{5}N](C^{-1}\slashed{P}')_{\alpha'\beta'}(\slashed{k}_{2}-m_{2})_{\alpha'\alpha}(\slashed{k}_{3}-m_{3})_{\beta'\beta}\iu(\sigma_{\sigma\rho}P^{\rho}C)_{\alpha\beta}\mathcal{T}^{*}\mathcal{V}\nn\\
     &+ [\overline{N}^{'}\g_{5}\g_{\mu_{2}}(\slashed{k}_{3}+m_{3})\g^{\mu_{1}}\g_{5}N]\iu (C^{-1}\sigma^{\nu_{2}\mu_{2}}P_{\nu_{2}}^{'})_{\alpha^{'}\beta^{'}}\g^{\mu}_{\alpha^{'}\alpha}(\slashed{k}_{2}-m_{2})_{\beta^{'}\beta}\iu(\sigma_{\mu_{1}\nu_{1}}P_{\nu_{1}}C)_{\alpha\beta} \nn
     \mathcal{T}^{*}\mathcal{T}
\\
     &+ [\overline{N}^{'}\g_{5}\g_{\mu_{2}}(\slashed{k}_{3}+m_{3})\g^{\mu_{1}}\g_{5}N]\iu (C^{-1}\sigma^{\nu_{2}\mu_{2}}P_{\nu_{2}}^{'})_{\alpha^{'}\beta^{'}}\g^{\mu}_{\beta^{'}\beta}(\slashed{k}_{2}-m_{2})_{\alpha^{'}\alpha}\iu(\sigma_{\mu_{1}\nu_{1}}P_{\nu_{1}}C)_{\alpha\beta} 
     \mathcal{T}^{*}\mathcal{T}
 \nn\\
     &+ \dots \big\}
\label{eq:aftdelta}
\ea
The $1/Q^{2}$ factor after the integration measure comes from the $Q$ dependence of
$\delta(k_{i}^{0}-k^{'0}_{i}) = \delta((k_{i}^{+}+x_{i}Q/\sqrt{2})-(k_{i}^{'+}+x'_{i}Q/\sqrt{2}))$.
Evaluating the first two terms in the above expression and isolating the $F_{2}$ contribution gives

\ba
& \overline{N}^{\prime}\frac{\iu}{2}\sigma^{\mu\nu}q_{\nu}N \, F_{2}  \sim \int dk_{1y}dk_{2y}dx_{1}dx_{2}\frac{1}{Q^2}\,\ 
\left[\overline{N}'\frac{\iu}{2m_{P}}\sigma^{\mu\nu}q_{\nu}N\right]\,\ 8m_{P}[(P\cdot k_{2})m_{3}+(P\cdot k_{3})m_{2}]\,\ \mathcal{T}^{*}\mathcal{V}
\nn \ea The other terms are similar.

\subsection{The endpoint wave function and $F_{2}$} 

 The leading power wave functions of Ref.\cite{Dagaonkar:2014} were determined in the endpoint region:
\ba \mathcal{V,A,T} \propto (1-x_{1})x_{1}\e^{- k_{T}^{2}/\Lambda^{2}}.
\label{eq:xdep} \ea
The exponential dependence on the transverse momentum is a generic form that restricts the range of $x_1\in(1-\frac{\Lambda}{Q},1)$ and $x_2\in(0,\frac{\Lambda}{Q})$.

The dot products are 
\ba
&P\cdot k_{i} = k^{0}_{i}\sqrt{\frac{Q^2}{2}+m^{2}_{p}}-x_{i}\frac{Q^{2}}{2}\,.\nn
\ea
In terms of the light cone variables, this gives
\ba
P\cdot k_{i}= & (k_{i}^{+}+x_{i}\frac{Q}{\sqrt{2}})\sqrt{\frac{Q^2}{2}+m^{2}_{p}}-x_{i}\frac{Q^{2}}{2}  \nn \\  & \qquad  \sim \left(\frac{m^{2}_{q_{i}}+k_{\perp i}^{2}}{x_{i}Q}+x_{i}\frac{Q}{\sqrt{2}}\right)\frac{Q}{\sqrt{2}}-x_{i}\frac{Q^{2}}{2}\sim \Lambda Q \nn
\ea

It follows that
\ba
F_{2} &\propto \int dk_{1y}dk_{2y}dx_{1}dx_{2}\frac{1}{Q^2} \nn \\ 
      &\hspace{15mm}\times 8m_{P} [(\Lambda Q) m_{3}+(\Lambda Q)m_{2}]\,\ (1-x_{1})\e^{-\frac{ k_{T}^{2}}{\Lambda^{2}}}
\,\ (1-x'_{1})\e^{-\frac{k_{T}^{'2}}{\Lambda^{2}}} \nn \\
& \hspace{15mm} \sim \int dx_{1}dx_{2}\frac{1}{Q^2} Q (1-x_{1})(1-x_{1})\e^{-\frac{ k_{T}^{2}}{\Lambda^{2}}}\e^{-\frac{k_{T}^{'2}}{\Lambda^{2}}} \nn \\
&  \hspace{30mm}= \frac{1}{Q^2} Q \frac{1}{Q^3}\frac{1}{Q} = \frac{1}{Q^{5}} \label{eq:F2final}
\ea

\subsection{The ratio of form factors}

In our estimate of the form factor $F_2$ we used the wave 
function given in Eq. \ref{eq:xdep}, whose $x$ dependence was determined by 
fitting the Dirac form factor, $F_1$. However it is easy to see that
the ratio $F_2/F_1$ is independent of the precise form of the wave function
within the end point model. 

Consider a rather arbitrary wave function
\ba \mathcal{V,A,T} \propto f(x_1)\e^{- k_{T}^{2}/\Lambda^{2}}.\label{eq:xdep_gen} \ea
This leads to the Dirac form factor \cite{Dagaonkar:2014},
\ba
F_{1} &\propto \int dk_{1y}dk_{2y}dx_{1}dx_{2}\frac{1}{Q^2} [ 8 Q^2 m_{2}m_{3}]\,\ f(x_1)\e^{- k_{T}^{2}/\Lambda^{2}}\,\ f(x'_1)  \e^{- k_{T}^{'2}/\Lambda^{2}} \label{eq:F1final}
\ea
Similarly the form factor $F_2$ becomes,  
\ba
F_{2} \propto \int dk_{1y}dk_{2y}dx_{1}dx_{2}\frac{1}{Q^2}  
       8m_{P} [(\Lambda Q) m_{3}+(\Lambda Q)m_{2}]\, 
f(x_1)\e^{- k_{T}^{2}/\Lambda^{2}}\,f(x'_1)e^{- k_{T}^{'2}/\Lambda^{2}}.
\label{eq:F2final1} \ea

Taking the ratio gives
\ba\frac{F_{2}}{F_{1}}\propto\frac {\frac{1}{Q^2} 8m_{P} [(\Lambda Q)m_{3}+(\Lambda Q)m_{2}]}
{\frac{1}{Q^2}[8 Q^2 m_{2}m_{3}]}\propto \frac{1}{Q}. 
\label{eq:ratioF2F1}
\ea
Thus the ratio of form factors in the endpoint model is independent of the precise form of the wave function.  

The JLAB data \cite{jlab-gayou} shows
$QF_{2}/F_{1} \sim$ constant starting from $Q^2$ as low as 2 GeV$^2$. At such
low values $F_1$ differs significantly from its high-$Q^2 $scaling behavior, which 
is observed to set in for $Q^2 > 5\ {\rm GeV}^2$\cite{Sill:1992}.  In the low $Q^2$ regime a more complicated wave function is needed to fit the data.  However Eq. \ref{eq:ratioF2F1} follows quite generally since the dependence
on the wave function cancels out while taking the ratio.

\section{Soft gluon exchange}
\begin{figure}[!t]
\centering
\includegraphics[scale=0.75,angle=0]{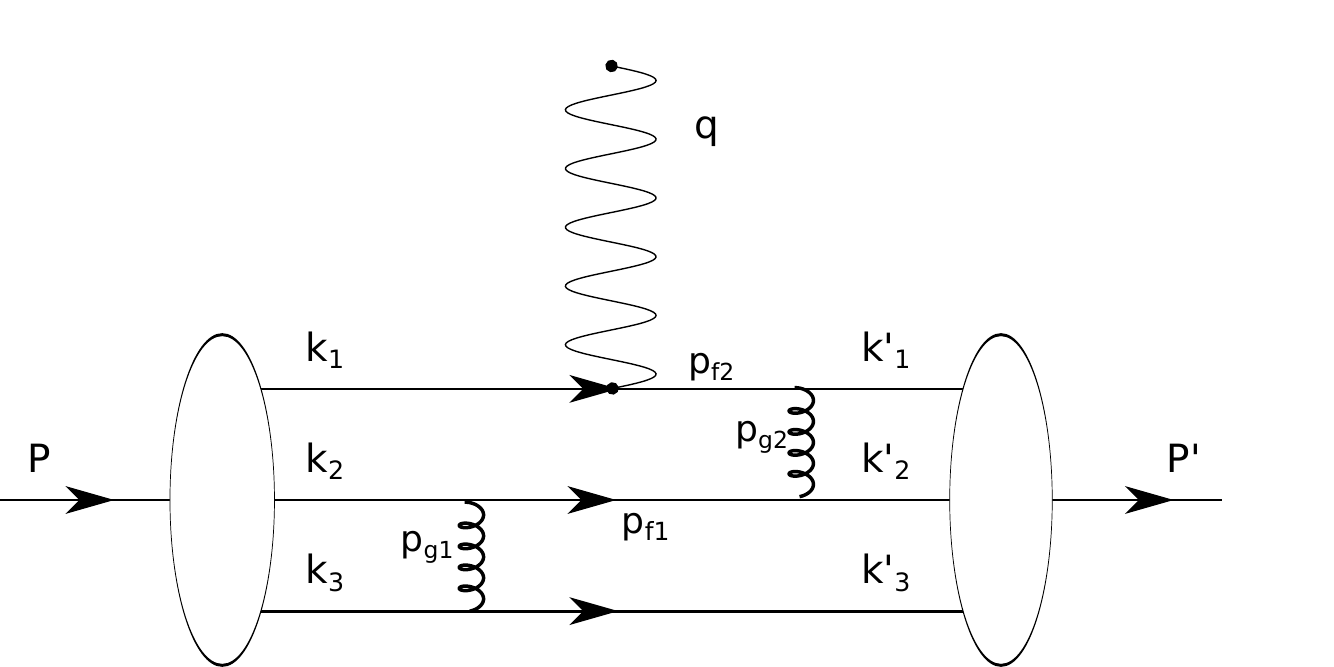}
\caption {\small A 2 gluon exchange contribution to the proton form factor
}
\label{fig:2gluon}
\end{figure}

It can be verified that addition of low momentum gluons in the interaction will not change the scaling behavior of the 
Pauli Form factor $F_{2}$. Consider the simple case of 2 gluon exchange illustrated in Fig.[\ref{fig:2gluon}].

The matrix element for this diagram is
\ba
&\nn \int[dk_{i}][dk'_{i}]  \,\ \bigg[\iu(C^{-1}\sigma^{\nu\sigma}P'_{\nu})_{\alpha'\beta'}(\overline{N}\g_{5}\g_{\sigma})_{\g'}\mathcal{T}^{*}\bigg]\bigg[[(-\iu g_{s}\g^{\rho})\frac{\iu(\slashed{p}_{f_{1}}+m_{1})}{p_{f_{1}}^{2}-m_{1}^{2}}(-\iu e \g^{\mu})]_{\g'\g}\frac{-\iu}{p_{g_{1}}^{2}}\frac{-\iu}{p_{g_{2}}^{2}}[-ig_{s}\g_{\lambda}]_{\beta'\beta}\\
&\hspace{3cm}[(-\iu g_{s}\g_{\rho})\frac{\iu(\slashed{p}_{f_{2}}+m_{1})}{p_{f_{2}}^{2}-m_{2}^{2}}(-\iu g_{s} \g^{\lambda})]_{\alpha'\alpha}\bigg]\bigg[(\slashed{P}C)_{\alpha\beta}(\g_{5}N)_{\g}\mathcal{V}\bigg]
\nn \ea
where
\ba
[dk_{i}] =\prod_{i}dx_{i}dk_{i\perp} \delta(x_{1}+x_{2}+x_{3}-1)\delta^{2}(\vec{k}_{1\perp}+\vec{k}_{2\perp}+\vec{k}_{3\perp})
\nn \ea
Evaluating the traces and extracting the co-efficient of $\overline{N}'\iu\sigma^{\mu\nu} q_{\nu}N$
we find
\ba
\overline{N}'\iu\sigma^{\mu\nu} q_{\nu}N \,F_{2} \sim \int [dk_{i}][dk'_{i}]g_{s}^{4}e\frac{8\iu m_{2}(p_{f_{1}}\cdot P)(\overline{N}'\iu\sigma^{\mu\nu} q_{\nu}N)\mathcal{T}^{*}\mathcal{V}}{(p_{f_{1}}^{2}-m_{1}^{2})(p_{f_{2}}^{2}-m_{2}^{2})p_{g_{1}}^{2}p_{g_{2}}^{2}}
\nn \ea
Keeping only leading power term for the limit $Q\gg \Lambda$, dropping transverse momentum integrals of order the hadronic scale and substituting $\mathcal{V},\mathcal{T}$ from Eq.\ref{eq:xdep} gives
\ba
F_{2}\sim \int dx_{1}dx_{2}dx'_{1}dx'_{2} \,\ \frac{8m_{2}g_{s}^{4}\frac{Q^{2}}{2}}{\Lambda^{2}(-(1-x_{1})Q^{2})\Lambda^{2}\Lambda^{2}}\times (1-x_{1})(1-x'_{1})
\nn \ea

Each integral $dx$ over an interval of length $\Lambda/Q$ contributes a power of $1/Q$. The
integration of $1-x$ over $1-\Lambda/Q< x<1$ contributes a power of $1/Q$. It follows that 
\ba
F_{2} \propto \int dx_{1}dx'dx_{2}dx'_{2}(1-x')\propto \frac{1}{Q\times Q \times...Q} \sim \frac{1}{Q^{5}}.
\nn \ea

Thus the gluon exchanges do not change the leading power behavior.

\section{Conclusions}

As mentioned in the Introduction, if the $EP$ model had been given adequate attention 30 years ago,
a fit to the known $1/Q^{4}$ dependence of $F_{1}$ would have then predicted $F_{2}/F_{1} \sim 1/Q$
at large $Q$, just as eventually observed. The calculation was never done, despite the model's
visibility after initial development by Drell, Yan, Feynman, and
others.\cite{Drell70,Feynman69,West70}. 

Between then and now came a period attempting to dispense with hadron structure in form factors, and
replacing protons with perturbation theory, which revealed very little about hadron structure. We
find that one simple pattern of an endpoint wave function, previously determined in Ref.
\cite{Dagaonkar:2014} and going like $1-x$, explains many independent experiments. The endpoint
region produces the original and earliest quark-counting model\cite{Drell70}. For each spectator
integration $dx$ restricted to $x \lesssim \Lambda/Q$ an integral goes like $\Lambda/Q$. For each
hard struck quark with $1-\Lambda/Q \lesssim x \leq 1$ an integral goes like $\Lambda/Q$. Thus three
quarks leads to $F_{1}\sim 1/Q^{4}$. The leading twist Dirac structure, which has no room for
orbital angular momentum, still allows a reversal of the proton's chirality characterizing $F_{2}$,
and $F_{2}\sim 1/Q^{5}$. These are not asymptotic limits, but generic results of power-counting that
apply in the region $Q >> \Lambda$, namely $Q \gtrsim $ GeV.

The fact that $QF_{2}(Q^{2})/F_{1}(Q^{2})$ is nearly constant with $Q$ down to rather low values of
momentum transfer is now understood. At small $Q$ the details of the endpoint wave function enter
the calculation, and replacing $dx \sim \Lambda/Q$ is not accurate. It is possible to fit that
dependence from data for $F_{1}$ rather trivially. However the integrations for $F_{2}$ are so
nearly like those for $F_{1}$ that the details of the wave function cancel out in the ratio
$F_{2}/F_{1}$. The rule that $F_{2}/F_{1}\sim 1/Q$ for $Q>>$GeV naturally extends itself into the
region of $Q\sim $ few GeV. When future experiments probe higher momentum transfers we are confident
that $QF_{2}(Q^{2})/F_{1}(Q^{2})$ will remain constant, regardless of what might occur with the
numerator and denominator.

\end{document}